\documentclass[11pt,twoside]{article}
\usepackage[pdftex]{graphicx}
\usepackage{amsmath}
\usepackage{amssymb}
\usepackage{cite}
\usepackage{amstext,amsthm}
\usepackage{graphicx}
\usepackage{epstopdf}
\usepackage{color}
\usepackage{bm}
\usepackage{appendix}
\usepackage[T1]{fontenc}
\usepackage{bbm}
\usepackage{latexsym}

\usepackage{float}
\usepackage{verbatim}
\usepackage{lipsum}
\usepackage{braket}

\begin{document}
\newcommand{\pst}{\hspace*{1.5em}}

\newcommand{\rigmark}{\em Journal of Russian Laser Research}
\newcommand{\lemark}{\em Volume 30, Number 5, 2009}

\newcommand{\be}{\begin{equation}}
\newcommand{\ee}{\end{equation}}
\newcommand{\ds}{\displaystyle}
\newcommand{\bea}{\begin{eqnarray}}
\newcommand{\eea}{\end{eqnarray}}
\newcommand{\ba}{\begin{array}}
\newcommand{\ea}{\end{array}}
\newcommand{\arcsinh}{\mathop{\rm arcsinh}\nolimits}
\newcommand{\arctanh}{\mathop{\rm arctanh}\nolimits}
\newcommand{\bc}{\begin{center}}
\newcommand{\ec}{\end{center}}

\thispagestyle{plain}

\label{sh}


\begin{center} {\Large \bf
\begin{tabular}{c}
STRATEGIES FOR VARIATIONAL QUANTUM COMPILING OF A \\[-1mm]
ZERO-PHASE BEAMSPLITTER ON THE XANADU X8 PROCESSOR
\end{tabular}
 } \end{center}

\bigskip

\bigskip

\begin{center} {\bf
T.J. Volkoff
}\end{center}

\medskip

\begin{center}
{\it
Theoretical Division, Los Alamos National Laboratory,\\
Los Alamos, NM, USA.

\smallskip

$^*$Corresponding author e-mail:~~~volkoff@lanl.gov\\
}\end{center}

\begin{abstract}\noindent
In the context of variational compiling of a continuous-variable (CV) unitary operation, the architecture and parameter space of the Xanadu X8 processor constrain both the set of feasible compiling problems and the allowed cost functions. In this paper, we motivate a faithful cost function for variational compiling of a two-mode, continuous-variable beamsplitter gate with real matrix elements (i.e., a ``zero-phase'' beamsplitter) that complies with the constraints of the X8 processor. This cost function is then computed on the X8. Despite the noise in the processor, we find that the cost function exhibits optimum parameter resilience and, therefore, that this variational compiling problem is feasible on the X8.  The intent of the paper is partly to report a proof-of-principle cost function calculation on near-term CV hardware, and partly to present methods that may be relevant for CV variational compiling problems in more complicated, large scale settings.
\end{abstract}

\medskip


\section{Introduction}\label{sc:intro}
\begin{figure}[t]
    \centering
    \includegraphics[scale=1.3]{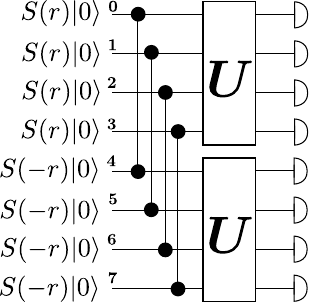}
    \caption{Architecture of the X8, $r=1$. The squeezed mode pairs $(0,4), \ldots, (3,7)$ are coupled by beamsplitter to produce $\ket{\text{TMSS}_{r=1}}$ in each mode pair. The same unitary $U$ must be applied to registers $(0,1,2,3)$ and $(4,5,6,7)$.}
    \label{fig:x8c}
\end{figure}

Several simple variational quantum algorithms for the task of compiling continuous-variable (CV) quantum gates were developed in \cite{volk}, and a detailed background, motivation, and list of references related to the subject of variational quantum compiling can be found in that paper. One of the algorithms proposed in \cite{volk} utilizes CV entanglement in the form of two-mode squeezed states. Two-mode squeezed states are now accessible by a remote user for cloud quantum computing applications. Specifically, the Xanadu X8 processor (hereafter ``X8'') is a photonic chip (10 mm $\times$ 4 mm chip embedded with silicon nitride waveguides) coupled to a high-rate photon number resolving (PNR) detector (four-photon detection at $10^{4}$ counts per second, on average). More details of the device and some potential applications can be found in \cite{arra}. Fig. 1a of \cite{arra} shows the X8 circuit architecture, and this has been adapted to the notation of the present work in Fig. \ref{fig:x8c}. Although the architecture constraints of the X8 are incompatible with solving universal CV variational quantum compiling problems; in this work, we will demonstrate that they admit the possibility of solving a simple CV variational quantum compiling problem. 

A general definition of a universal variational quantum compiling problem, and a specific example in the CV setting is provided in Section \ref{sec:b}. Restricting further to the subset of CV variational quantum compiling problems that are feasible on the X8, we consider in Section \ref{sec:costmod} how to use the X8 to compute a cost function $C(x)$ that depends on a real parameter $x$ through a two-mode, linear optical unitary operation $V(x)$, and possesses a global minimum $x_{c}$ such that the unitary $V(x_{c})$ is equal to a target unitary $U$.   Specifically, the cost function in the central example in Section \ref{sec:costmod} depends on $x$ through the unitary 50:50 beamsplitter operation $V(x)=e^{{\pi\over 4}(e^{ix}a^{\dagger}b - e^{-ix}b^{\dagger}a)}$, $x\in [-\pi,\pi)$, and the unique global minimum of the cost function is attained at $x_{c}=0$, i.e., the target unitary $U=V(x_{c})$ is a zero-phase, 50:50 beamsplitter (and therefore a real matrix in the Fock basis). In Section \ref{sec:iii}, we perform a noiseless numerical simulation of the cost function for compiling the zero-phase, 50:50 beamsplitter. This cost function depends on single PNR measurement outcome.  Our main result (Section \ref{sec:cfcalc}) is that this cost function exhibits strong optimal parameter resilience \cite{opr} to the noise in the X8. This means that despite the noise in the X8, the unique global minimum $x_{c}=0$ of the cost function is the also the global minimum of the cost function estimate obtained from X8 PNR measurement data. This result shows that the unique global minimum can be identified to a precision limited by device losses and the number of PNR measurement shots. Therefore, by coupling the X8 processor to a classical optimization algorithm, it is possible in principle to implement a variational quantum compiling algorithm on this device. 

Beyond the simple task of compiling a zero-phase beamsplitter, the results of this paper suggest that compiling arbitrary linear optical unitaries is feasible on a hypothetical, less constrained version of the X8 that allowed the application of different four-mode linear optical unitaries on the entangled two-mode squeezed registers. The results further demonstrate some practical methods for obtaining cost function estimates based on photon counting measurement. In the present work, we focus on motivating a faithful cost function for variational quantum compiling on a one-dimensional submanifold of linear optical unitaries and computing it on the X8. A full implementation of variational quantum compiling is a hybrid quantum-classical algorithm in which estimates of a cost function are obtained from the quantum module of the algorithm and the classical module carries out the optimization. The present work focuses solely on the quantum module, viz., obtaining the best possible cost function estimates from the X8.

\section{Background\label{sec:b}}
Before discussing the main result, we presently review the definition of a universal variational quantum compiling problem \cite{volk} and discuss why such a general framework cannot be used to carry out a proof-of-principle CV variational quantum compiling experiment on the X8. The ingredients of a universal quantum compiling problem (also known as a \textit{quantum circuit synthesis} problem) on $n$ modes are: 1. the free group $F(\mathcal{W})$ where $\mathcal{W}$ is the union of $K$ submanifolds of the unitary group on $n$ modes (can write an element of each submanifold as $W_{k}(\alpha^{(k)})$ where $k=1,\ldots, K$ and $\alpha^{(k)}\in \mathbb{R}^{m(k)}$ is the $m(k)$-dimensional real parametrization), 2. a target unitary operation $U$ which can be any unitary operator, 3. a cost function $C: \lbrace U\rbrace \times F(\mathcal{W}) \rightarrow [0,1]$ with global minimum at $(U,W_{\text{opt}})$ such that $W_{\text{opt}}$ also minimizes $\Vert W-U\Vert$ over $F(\mathcal{W})$ (or a constrained subset of $F(\mathcal{W})$). The index $k$ corresponds to the gate type and subset of registers on which $W_{k}(\alpha^{(k)})$ acts. For discrete variable quantum circuits, the Solovay-Kitaev theorem can be formulated as a set of sufficient conditions for universal quantum compiling with respect to the operator norm. Other kinds of cost functions in the quantum compiling problem allow a quantum compiling algorithm to be efficiently carried out on a quantum computer (e.g., via the SWAP test or its CV version \cite{vs}). The full variational quantum compiling algorithm is a hybrid quantum-classical algorithm in which $C$ is estimated from measurement of a quantum circuit (viz., on the quantum computer), then the element of $F(\mathcal{W})$ is updated according to a classical optimization algorithm which takes as input the previous estimates of $C$. When $K=1$, optimization of the cost function $C$ is just over the coordinates $\alpha^{(1)}$ of $W_{1}(\alpha^{(1)})$. This situation occurs, e.g., in the problem of fixed-gate compiling, in which the quantum circuit structure is fixed during the optimization, leaving only an optimization over the continuous coordinates. This is the problem that we consider in the present work.  Because any two-mode linear optical unitary can be implemented on the X8 by specifying the appropriate parameters in its rectangular decomposition \cite{rect}, any feasible variational quantum compiling problem on the X8 could be formulated as fixed-gate compiling. Throughout this work, a cost function $C$ is called \textit{faithful} if for any target unitary $U$, the unique global minimum of $C$ is obtained at $W$ such that $W=U$.

We now illustrate a universal variational quantum compiling problem in the CV setting. Let $\ket{\text{TMSS}_{r}}_{AB}:= \otimes_{j=1}^{M}\ket{\text{TMSS}_{r}}_{A_{j}B_{j}}$ be a two-mode squeezed state on registers $A=(A_{1},\ldots, A_{M})$, $B=(B_{1},\ldots, B_{M})$ with $A_{j}$ and $B_{j}$ a single CV mode \cite{agarwal}. Its Fock basis amplitudes are obtained from $\left( \ket{n}_{A_{j}}\otimes \ket{m}_{B_{j}},\vert\text{TMSS}_{r}\rangle_{A_{j}B_{j}}\right) = \delta_{n,m}{\tanh^{n}r\over \cosh r}$. A cost function for compiling an $M$-mode unitary $U$ from a parameterized $M$-mode unitary $V(x)$ ($x\in \mathbb{R}^{n}$ are the parameters) is given by
\begin{equation}
C(x)=1-\vert \langle \text{TMSS}_{r} \vert_{AB} U_{A}\overline{V(x)}_{B} \vert \text{TMSS}_{r}\rangle_{AB}\vert^{2}
\label{eqn:tmss}
\end{equation}
where we have shortened the cost function notation from the burdensome expression $C(U,V(x))$.
The parameterized ansatz $V(x)$ is called \textit{expressive} if there exists $x$ such that $V(x)=U$. In this paper, we only utilize expressive ansatze. It is not difficult to show that the global minimum of the cost function (\ref{eqn:tmss}) approaches the value 0 if and only if $V(x)$ is expressive and $r\rightarrow \infty$, with the global minimum being obtained at $x_{c}$ such that $V(x_{c})=U$. Therefore, to compile a unitary $U$ by minimization of cost function (\ref{eqn:tmss}), it is sufficient to have access to a two-mode squeezed state of sufficiently large squeezing and an expressive ansatz $V(x)$. Two measurement schemes that yield (\ref{eqn:tmss}) are: 1. application of a post-processing quantum circuit that inverts the two-mode squeezing, followed by a PNR measurement to obtain the vacuum probability, or 2. implementation of a subvacuum noise Gaussian measurement corresponding to the squeezing value $r$. Unfortunately, regardless of the measurement scheme, such a universal CV variational compiling algorithm cannot be implemented on the X8 because if $U$ is not a linear optical unitary, it is impossible to construct an expressive ansatz $V(x)$ using the linear optical circuit available on the X8 device. Further, the X8 requires that the unitary operators are the same on the registers $A$ and $B$. In Section \ref{sec:costmod} we leave behind the setting of universal variational compiling and and focus on designing a variational compiling problem with a cost function that can be computed on the X8.

\section{Cost function modification for the X8\label{sec:costmod}}
To calculate the cost function (\ref{eqn:tmss}) on a CV processor in a minimum setting, i.e., compiling a single-mode unitary $U$, it is clear that only two modes are needed ($M=1$). Further, in many cases, it has been found that squeezing with $r=O(1)$ is sufficient for minimization of (\ref{eqn:tmss}) for few-mode variational CV quantum compiling. Therefore, the capabilities of the X8 for proof-of-principle variational quantum compiling are not limited by squeezing strength or number of modes. Even so, computation of (\ref{eqn:tmss}) on the X8 is infeasible for a general unitary $U$ on say, four modes, due to the circuit architecture and unitary manifold constraints of the X8. We enumerate these constraints as follows \cite{xan}:
\begin{enumerate}
\item Labeling the X8 modes by $0,\ldots, 7$, the squeezing parameter $r\in \lbrace 0,1\rbrace$ and the squeezing connectivity is 04, 15, 26, 37.
\item  Squeezing occurs at beginning of the circuit (no in-line squeezing). The measurement is PNR measurement.
\item The unitary loaded onto the $0123$ and $4567$ modes must be the same, and this unitary must be in the subgroup generated by phase shifts and beamsplitters (linear optical unitaries).
\end{enumerate} These constraints make it clear that (\ref{eqn:tmss}) cannot in general be calculated on the X8 because the $A_{1}A_{2}A_{3}A_{4}$ ($B_{1}B_{2}B_{3}B_{4}$) modes correspond respectively to $0123$ ($4567$) and the unitaries $U_{A}$ and $V(x)_{B}$ are in general distinct (not to mention the fact that one could be interested in compiling a non-linear optical operation). However, because a two-mode squeezed state is the input for both the cost function (\ref{eqn:tmss}) and the X8 processor, the form of (\ref{eqn:tmss}) can be used as a template to define a valid variational compiling problem on the X8.

 When the $A$ and $B$ registers carry the same unitary, the asymptotic ricochet property of the two-mode squeezed state is
\begin{equation}
\lim_{r\rightarrow \infty}V_{A}\otimes V_{B}\ket{\text{TMSS}_{r}}_{AB} = \lim_{r\rightarrow \infty} V_{A}V^{T}_{A}\otimes \mathbb{I}_{B}\ket{\text{TMSS}_{r}}_{AB}.
\label{eqn:ric}
\end{equation}
If $V(x)$ is an ansatz such that $V(x)$ is real for a single value $x=x_{c}$, it then follows that a modified version of cost function (\ref{eqn:tmss}) of the form
\begin{equation}
C_{2}(x)=1-\vert \langle \text{TMSS}_{r}\vert_{AB}V(x)_{A}\otimes V(x)_{B}\vert \text{TMSS}_{r}\rangle_{AB}\vert^{2}
\label{eqn:realtmss}
\end{equation}
approximately obtains its global minimum value 0 at a real unitary $V(x_{c})$ for sufficiently large $r$ (see Appendix A of \cite{volk} for details of the approximation). The restricted task of compiling a real unitary allows to remove the fiducial unitary $U=V(x_{c})$ from the cost function (\ref{eqn:tmss}) and load the ansatz $V(x)$ on both $A$ and $B$ registers, while remaining a faithful cost function as long as $x_{c}$ is the unique global minimum. The requirement that $V(x)$ be real at a unique point can be loosened by allowing a discrete set of real points of $V(x)$ and restricting the optimization to disjoint neighborhoods of the points in that set. If $V(x)$ is a linear optical unitary, the cost function $C_{2}$ is consistent with constraints 1. and 3. of the X8. However, it is not compatible with constraint 2., and different cost functions descending from (\ref{eqn:realtmss}) can be introduced case-by-case. We now provide examples of such cost functions for three linear optical unitary variational compiling problems. In all examples, we will need the Fock probabilities $p(\vec{n}_{AB},r,V):= \vert \langle \vec{n}\vert V_{A}\otimes V_{B} \ket{\text{TMSS}_{r}}_{AB}\vert^{2}$, $\vec{n}_{AB}\in (\mathbb{Z}_{\ge 0})^{\times 2M}$, that occur when a unitary $V$ is loaded into both the $A$ and $B$ registers.

\textbf{Example 1 (phase space reflection)}: We start with a deceptively simple example of compiling a phase space reflection ($A=A_{1}$, $B=B_{1}$). In this example, our aim of constructing a faithful cost function from the Fock probabilities $p(\vec{n}_{AB},r,V)$ will fail for a simple reason that we discuss below. We resolve this issue and construct an alternative cost function, but the resolution depends on successfully carrying out Example 2, which is the main focus of the paper.

Note that a phase shift $V(\phi)_{A}=e^{ i\phi a_{A_{1}}^{\dagger}a_{A_{1}}}$ is real if and only if $\phi\in \lbrace 0 ,\pi\rbrace$.  The property 
\begin{equation}
V(\phi)_{A_{1}}\otimes V(\phi)_{B_{1}}\ket{\text{TMSS}_{r}}_{A_{1}B_{1}}=\ket{\text{TMSS}_{r}}_{AB}
\label{eqn:nnn}
\end{equation}(which is stronger than (\ref{eqn:ric})) holds for all $r$ if and only if $\phi\in \lbrace 0 ,\pi\rbrace$, i.e., if $V(\phi)$ is a real unitary. Naively, it may now seem that one can faithfully compile the phase space reflection $e^{i\pi a^{\dagger}_{A_{1}}a_{A_{1}}}$, on the X8 processor by utilizing a cost function that compares $p(\vec{n}_{A_{1}B_{1}},1,V(\phi))$ on the interval $[{\pi\over 2},{3\pi\over 2}]$ to $p(\vec{n}_{A_{1}B_{1}},1,\mathbb{I})$. Indeed, when first considering the possibility, the author further noted that instead of the computing these probabilities serially on the two mode register $A_{1}B_{1}=04$, the X8 connectivity permits parallel computation of these probabilities using only four modes (e.g., on $A_{1}A_{2}B_{1}B_{2}=0145$): simply load $V(\phi)_{A}$ on  $A_{1}=0$, $V(\phi)_{B}$ on $B_{1}=4$ as before, but compute the fiducial probability $p(\vec{n}_{A_{2}B_{2}},1,\mathbb{I})$ instead of the mathematically equivalent $p(\vec{n}_{A_{1}B_{1}},1,\mathbb{I})$ (one could take $A_{2}B_{2}=15$ on the X8).   This parallelization would reduce the total number of shots required to get a high-quality cost function estimate. Despite this possibility, one must step back and note the fundamental point that the function $p(\vec{n}_{A_{1}B_{1}},1,V(\phi))$ is actually independent of $\phi$ because phase space rotations cannot be sensed by using a readout that has no coherence in the Fock basis. To introduce a viable cost function, one could use an estimate of one of the probabilities from a Bell measurement in the four dimensional subspace spanned by ``computational basis'' $\lbrace \ket{0}_{A_{1}}\ket{0}_{B_{1}},\ket{1}_{A_{1}}\ket{0}_{B_{1}},\ket{0}_{A_{1}}\ket{1}_{B_{1}},\ket{1}_{A_{1}}\ket{1}_{B_{1}}\rbrace$. Specifically, consider the probabilities 
\begin{align}
&{}p\left( (0,1,0,1)_{A_{1}A_{2}B_{1}B_{2}},r,U_{\text{BS}}({\pi\over 4},0)V(\phi)\right)\nonumber \\
&={\tanh^{2}r\over \cosh^{4}r}\cos^{2}{\phi\over 2}\nonumber \\
& p\left( (0,1,0,1)_{A_{3}A_{4}B_{3}B_{4}},r,\mathbb{I})\right)\nonumber \\
&={\tanh^{2}r\over \cosh^{4}r}
\label{eqn:reflpr}
\end{align}
where $U_{\text{BS}}({\pi\over 4},0)_{A}=e^{{\pi\over 4} (a_{A_{1}}^{\dagger}a_{A_{2}}-h.c.)}$ is a zero-phase 50:50 beamsplitter. The quotient of the first probability and the second probability is $\cos^{2}{\phi\over 2}$ which on $[{\pi \over 2},{3\pi\over 2}]$ gives an $r$-independent, faithful cost function for compiling the phase space reflection. The probabilities (\ref{eqn:reflpr}) can be computed in parallel on the X8, but computation of the first probability requires access to a zero-phase 50:50 beamsplitter at the end of the circuit. 
In the next example, which is the focus of the present work, we will construct a cost function for compiling a zero-phase 50:50 beamsplitter in terms of PNR measurement outcomes without any extra layers. Example 3 (see below), which involves compiling the transmissivity parameter of a beamsplitter,  also makes use of a zero-phase 50:50 beamsplitter. Therefore, one can view the next example as the ``central task'' for variationally compiling a two-mode linear optical unitary. 
 
\textbf{Example 2 (zero-phase beamsplitter)}: Consider the fact that a beamsplitter  $V(\theta,\phi)_{A}=e^{\theta (e^{i\phi}a_{A_{1}}^{\dagger}a_{A_{2}}-h.c.)}$ with fixed $\theta
\in (0,\pi]$ is real if and only if $\phi=0$ or $\phi=\pi$.  We will sometimes use the notation $U_{\text{BS}}( \theta,\phi)$ for $V(\theta,\phi)$, noting that it can be implemented using the \texttt{BSgate} function from StrawberryFields (StrawberryFields is the application programming interface (API) developed by Xanadu that allows users to access many useful functions from the theory of quantum optics, and compute the compatible ones on the X8). Again, the property (\ref{eqn:nnn}) holds for all $r$ if and only if $\phi\in \lbrace 0 ,\pi\rbrace$ (i.e., the beamsplitter is real, i.e., zero-phase or $\pi$-phase). From these facts and the list of constraints of the X8, it follows that one should be able to faithfully compile the real unitary $V(\theta,0)_{A}=e^{\theta (a_{A_{1}}^{\dagger}a_{A_{2}}-h.c.)}$ on the X8 processor by utilizing a cost function that compares $p(\vec{n}_{A_{1}A_{2}B_{1}B_{2}},1,V(\theta,\phi))$ on the interval $[-{\pi\over 2},{\pi\over 2}]$ to $p(\vec{n}_{A_{1}A_{2}B_{1}B_{2}},1,\mathbb{I})$.  As in Example 1, the X8 connectivity permits parallel computation of these probabilities using all 8 modes: just relabel the register for the second probability via $A_{1}A_{2}B_{1}B_{2}\mapsto A_{3}A_{4}B_{3}B_{4}$ and let $A_{3}A_{4}B_{3}B_{4}=2367$ on the X8. This is the example that we focus on in the rest of the paper, although when we actually compute the cost function on the X8 in Section \ref{sec:cfcalc}, we will find that the cost function quality is improved if these probabilities are computed serially. To make a connection to the circuit diagram in \cite{xan}, note that $U=V(\theta,\phi)_{01} \otimes \mathbb{I}_{23}$.

The question remains how to construct a cost function that compares $p(\vec{n}_{A_{1}A_{2}B_{1}B_{2}},1,V(\theta,\phi))$ and $p(\vec{n}_{A_{3}A_{4}B_{3}B_{4}},1,\mathbb{I})$. Let us fix $\theta = {\pi\over 4}$; any other nonzero value of $\theta$ is also acceptable. If one intends to compute these in parallel on the X8, it is useful to relabel the respective probabilities as follows: \begin{equation}p(\vec{n}_{A_{1}A_{2}B_{1}B_{2}},1,V({\pi\over 4},\phi)) = q(\vec{n}_{0145},\phi)\; , \; p(\vec{n}_{A_{3}A_{4}B_{3}B_{3}},1,\mathbb{I})=p(\vec{n}_{2367}).\nonumber \end{equation}  Because we are comparing the photon number distributions linear optical unitaries acting on two-mode squeezed states, only a single $\vec{n}$ is necessary to distinguish the probability distributions. For example, one could define the cost function as a function $D(\phi)$ that is defined by the contribution of the Fock state $\ket{0,1,0,1}_{0415}$ to the total variation distance between photon number distributions:
\begin{align}
D_{1}(\phi)&=\vert q((0,1,0,1)_{0145},\phi) - p((0,1,0,1)_{2367}) \vert \nonumber \\
&= {\tanh^{2}r \over 2\cosh^{4}r}(1-\cos2\phi) \big\vert_{r=1}.
\end{align}
In contrast to Example 1, the 50:50 beamsplitter in $V({\pi\over 4},\phi)$ generates quantum coherence in the Fock basis and allows the Fock probabilities to be used in constructing a cost function. Cost function $D_{1}(\phi)$ is a fine choice for the value $r=1$ allowed in the X8 processor. However, for larger values of $r$, such a cost function goes to zero pointwise. When one has access to a large number of shots, this problem can be avoided by using the likelihood ratio to define the cost function $D_{2}(\phi)$, where
\begin{align}
D_{2}(\phi)&:=\vert 1 - {q((0,1,0,1)_{0145},\phi)\over p((0,1,0,1)_{2367})} \vert \nonumber \\
&={1\over 2}(1-\cos 2\phi).
\label{eqn:truecost}
\end{align}

Given $M$ shots on the X8, one defines the estimate $\hat{D}(\phi)$ of $D(\phi)$ through the empirical distributions \begin{equation}Q((0,1,0,1)_{0145},\phi):= {N((0,1,0,1)_{0145},\phi)\over M}\; , \; P((0,1,0,1)_{2367}):= {N((0,1,0,1)_{2367})\over M}\nonumber\end{equation} (symbol $N(\vec{n}_{X})$ stands for ``number of counts of $\vec{n}$ in register $X$'') by
\begin{equation}
\hat{D_{2}}(\phi)=\big\vert 1 - {Q((0,1,0,1)_{0145},\phi)\over P((0,1,0,1)_{2367})} \big\vert .
\label{eqn:costest}
\end{equation} 
Alternatively, one can define $N(\vec{n}_{X})$ as the number of times (over $M$ measurements) that the photodetectors in modes $X$ registered the pattern $\vec{n}$.
Note that $\lim_{M\rightarrow \infty}\hat{D_{2}}(\phi) =D_{2}(\phi)$ by law of large numbers. However, it should be noted that $q((0,1,0,1)_{0145},\phi)$ and $p((0,1,0,1)_{0145})$ decrease exponentially with increasing squeezing parameter $r>0$, so before using this cost function one should consider whether one has access to sufficiently many shots to get well-converged estimates $Q((0,1,0,1)_{0145},\phi)$ and $P((0,1,0,1)_{2367})$. Since $r=1$ on the X8, this is not an issue in the present work.  It is possible to use total occupations to construct a cost function, but one must increase the depth of the circuit (note that operations that change the phase parameter commute with all occupations). This requirement may run into a problem in that the added layers presume access to the unitary that one is trying to compile. Consider the fact that
\begin{align}
&{}\big\vert {\langle -ia_{0}^{\dagger}a_{1} +i a_{1}^{\dagger}a_{0} \rangle_{V({\pi\over 4},\phi)_{01}V({\pi\over 4},\phi)_{45}\ket{\text{TMSS}_{r}}_{04}\ket{\text{VAC}}_{15}}\over \langle a_{2}^{\dagger}a_{2} \rangle_{\ket{\text{TMSS}_{r}}_{26}\ket{\text{VAC}}_{37}}} \big\vert = \vert\sin\phi\vert 
\label{eqn:tr}
\end{align}
which is a well-defined cost function on $[-{\pi\over 2},{\pi\over 2}]$.
However, the numerator of (\ref{eqn:tr}) cannot be directly computed on the X8 because the operator $ -ia_{0}^{\dagger}a_{1} +i a_{1}^{\dagger}a_{0}$ cannot be measured. So one would have to compute it using the fact that 
\begin{equation} \langle -ia_{0}^{\dagger}a_{1} +i a_{1}^{\dagger}a_{0} \rangle_{\ket{\psi}}= \langle a_{0}^{\dagger}a_{0} - a_{1}^{\dagger}a_{1} \rangle_{V({\pi\over 4},0)_{01}\ket{\psi}},\end{equation}
which involves an extra beamsplitter layer. In fact, that extra beamsplitter layer $V({\pi\over 4},0)$ is precisely the unitary that we are trying to compile, and this is the reason why the cost function (\ref{eqn:tr}) is unsatisfactory for the present task.

\textbf{Example 3 (50:50 beamsplitter)}: It is also possible to compile the transmissivity parameter $\theta$ of $V(\theta,\phi)$ on the X8 architecture. Unlike the case of compiling the phase $\phi$, total occupations can be used to construct a cost function without increasing the depth of the circuit. For instance, a cost function for compiling $U=U_{\text{BS}}({\pi\over 4},0)$ using  $V(\theta,0)=U_{\text{BS}}(\theta,0)$ is given by
\begin{align}
\big\vert {\langle a_{0}^{\dagger}a_{0} -a_{1}^{\dagger}a_{1} \rangle_{V(\theta,0)_{01}V(\theta,0)_{45}\ket{\text{TMSS}_{r}}_{04}\ket{\text{VAC}}_{15}}\over \langle a_{2}^{\dagger}a_{2}  \rangle_{\ket{\text{TMSS}_{r}}_{26}\ket{\text{VAC}}_{37}}}\big\vert &=\vert\cos2\theta\vert
\end{align} 
which is minimized at $\theta={\pi\over 4}$ on $[0,\pi/2]$. 

\section{Simulating the cost function\label{sec:iii}}

The optimization of a cost function by a variational quantum algorithm involves updating the parameterized circuit $V(x)$ after each block of measurements giving a cost function estimate. If the cost function can be computed efficiently pointwise to small error, then the complexity of the optimization depends only on intrinsic features of the problem, i.e., features of the cost function landscape. Because the X8 processor is a near-term photonic device and we do not have direct access to the optical hardware on which it is based, we cannot simply deduce the sample complexity (i.e., number of photon detections) of computing the cost function (\ref{eqn:realtmss}) to some desired level of accuracy. Therefore, we focus on the computation of the cost function at each point in parameter space.

\begin{figure}[t]
    \centering
    \includegraphics[scale=0.6]{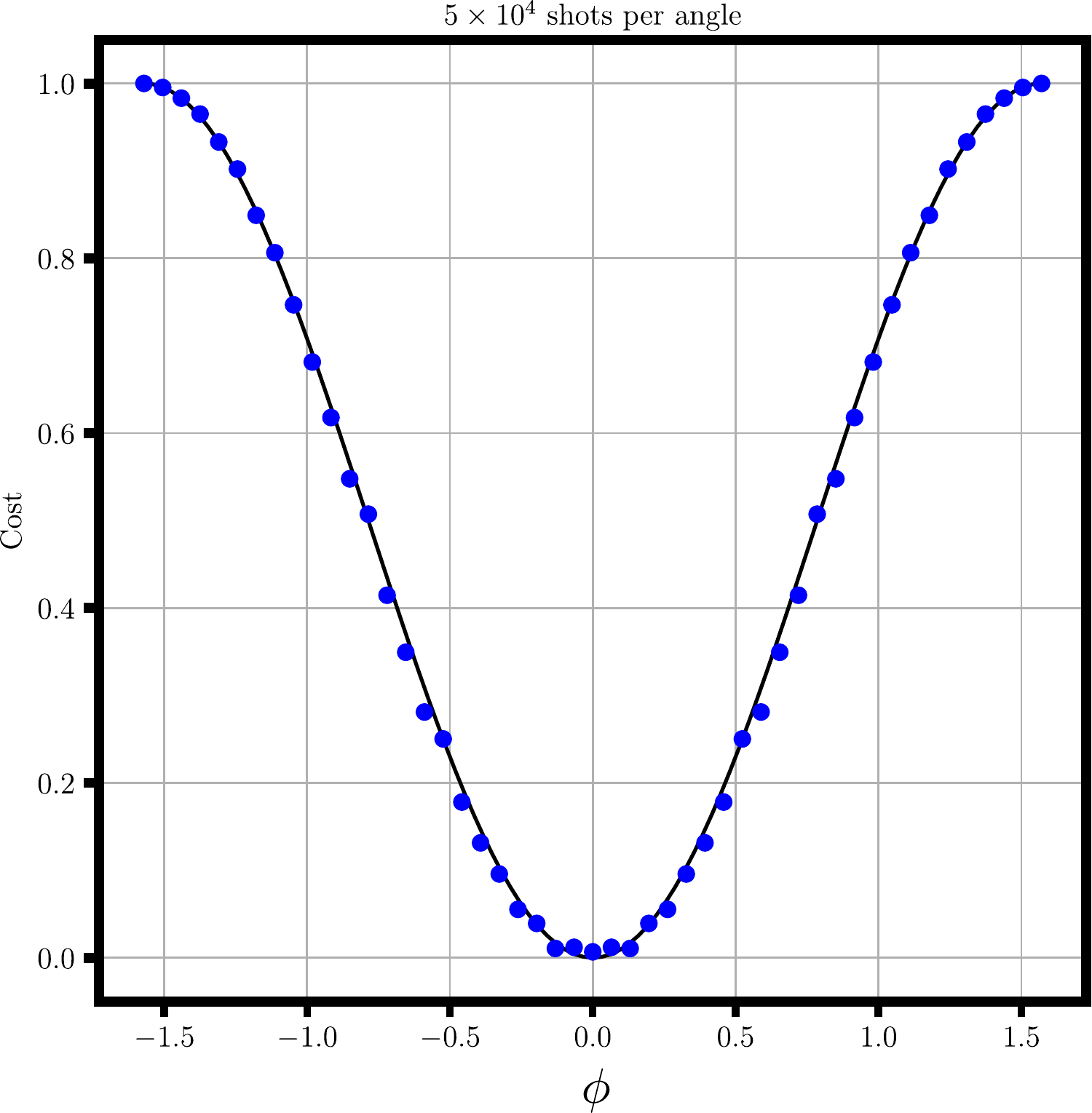}
    \caption{Cost function estimate (\ref{eqn:costest}) from simulated PNR measurement (blue dots) and analytical cost function (\ref{eqn:truecost}) (black). The cost function estimate is only computed on $[{\pi\over 2},0)$; the dots in $(0,{\pi\over 2}]$ are reflections about $\phi=0$.}
    \label{fig:rc}
\end{figure}

The StrawberryFields API provides useful functions to carry out numerical simulations of the X8 processor. For example, by using the TensorFlow backend with a certain Fock space cutoff, one can obtain the Fock basis (i.e., $\lbrace \otimes_{\ell=0}^{7}\ket{n_{\ell}} : n_{\ell}\in \mathbb{Z}_{\ge 0}\rbrace$) amplitudes for the state $V({\pi\over 4},\phi)_{01}V({\pi\over 4},\phi)_{45}\bigotimes_{X\in \lbrace 04,15,26,37\rbrace}\ket{\text{TMSS}_{1}}_{X}$ appearing in Example 2 above. Note that the PNR measurement of the X8 returns a vector in $\mathbb{N}_{\ge 0}^{\times 8}$ for every shot, so in reality the respective estimates of $q((0,1,0,1)_{0145},\phi)$ and $p((0,1,0,1)_{2367})$ are obtained as cumulative Fock basis probabilities from an 8-mode state. The Hilbert space dimension grows as the 8th power of the PNR measurement cutoff. However, this scaling can be circumvented in simulations by using the fact that the circuit is disconnected, and, for a given cutoff value, compute the probabilities $q((0,1,0,1)_{0145},\phi)$ and $p((0,1,0,1)_{2367})$ on separate 4-mode circuits as noncumulative probabilities. Using that method, one finds that a cutoff of 2 in the TensorFlow backend suffices for close agreement with the analytical cost function (see black curve of Fig. \ref{fig:rc}), and this approximate state-based simulation of the cost function can be done quickly.

A more realistic numerical simulation can be carried out by simulating a PNR measurement and using the cost function estimate (\ref{eqn:costest}). StrawberryFields provides this functionality with the \texttt{MeasureFock} operation. Again using the fact that the circuit is disconnected, the estimates $Q((0,1,0,1)_{0145},\phi)$ and $P((0,1,0,1)_{2367})$ are obtained separately. The numerical simulation (with number cutoff 10 and $M=5\times 10^{4}$ shots per $\phi$ value) is shown in Fig. \ref{fig:rc}.  One notices that there are no error bars on the cost function estimate. That is because the numerical simulation was not repeated. In a higher complexity scenario, one would repeat the simulation some number of times for each $\phi$ to get an accurate value of the cost pointwise. The appropriate number of times to repeat the experiment depends on intrinsic features of the parameter space, and computational features such as the total number of shots, so it would be worked out empirically according to the needs of the compiling algorithm. Also note that since $P((0,1,0,1)_{2367})$ is independent of $\phi$, it can be precalculated to some specified precision and then used as a parameter of the cost function, although that was not done for Fig. \ref{fig:rc}; both $Q((0,1,0,1)_{0145},\phi)$ and $P((0,1,0,1)_{2367})$ were estimated from $5\times 10^{4}$ photon number detections. Compared to obtaining the probabilities almost exactly from approximate calculation of the state vector, the estimates in the blue dots of Fig. \ref{fig:rc} are computationally expensive (about 5 minutes per $\phi$ value on an Intel i7-9850H at base clock rate 2.60 GHz).

The use of a single Fock amplitude $\ket{0,1,0,1}$ to define the cost function automatically implies a resolution/shot tradeoff when using PNR measurement to determine the cost function estimate. Consider the fact that
\begin{align}
&{}\vert \langle 0,1,0,1\vert_{0145} V({\pi\over 4},\phi)_{01} V({\pi\over 4},\phi)_{45} \bigotimes_{X\in \lbrace 04,15\rbrace}\ket{\text{TMSS}_{1}}_{X} \vert^{2} \nonumber \\ &{}\approx 0.05\cdot (1+\cos 2\phi).
\end{align}
At $\phi = \pm {\pi\over 2}$, the probability is 0. For $0<f\ll 1$, one needs to shift $\phi$ by about $\sqrt{10f}$ to boost the probability to $f$. The number of shots needed to resolve the probability $f$ is about $f^{-1}$. Therefore, having access to $M$ shots implies an uncertainty of ${\sqrt{10\over M}}$ in $\phi$ (this estimate does not take into account any circuit noise).

\section{Calculating the cost function on X8\label{sec:cfcalc}}

Unlike the numerical simulation of PNR measurement used in Section \ref{sec:iii} to obtain the estimate $\hat{D_{2}}(\phi)$, the X8 exhibits imperfections that can only be attributed to the nature of the device. These imperfections are evident even in trivial implementations of the device. For instance, loading a circuit with zero squeezing and identity linear optical transformation on the X8 results still results in accumulating some non-vacuum photon counts \cite{arra}.  In the best case scenario, these imperfections simply increase the sample complexity of variational compiling.

To improve the estimate (\ref{eqn:costest}) for a given number of X8 photon number measurements, it was found advantageous to calculate $P((0,1,0,1)_{0145})$ instead of $P((0,1,0,1)_{2367})$ (the estimate $Q((0,1,0,1)_{0145},\phi)$ is always computed on modes 0145). This doubles the number of photon measurements required for an estimate of (\ref{eqn:costest}) because it requires sequential access to the device. A hypothetical X8 device with identical two-mode squeezed mode pairs could enable higher quality parallel computation of the $(0,1,0,1)$ counts in (\ref{eqn:costest}).

Not surprisingly, no improvement was obtained by using only counts of the Fock state $\ket{0,1,0,1}_{0145}\ket{0,0,0,0}_{2367}$ to compute the estimates $Q$ and $P$, i.e., postselection on the vacuum in modes where there is no squeezing did not improve the cost function landscape. It is possible to define a cost function for this variational compiling problem that takes into account more measurement outcomes than $0101$. Such a cost function may have a lower sample complexity compared to the present cost function, which is based on the single outcome 0101. This point is especially relevant for CV quantum compiling at larger squeezing values, for which any given outcome occurs with probability approaching $0$.

\begin{figure*}[t]
    \centering
    \includegraphics[scale=0.5]{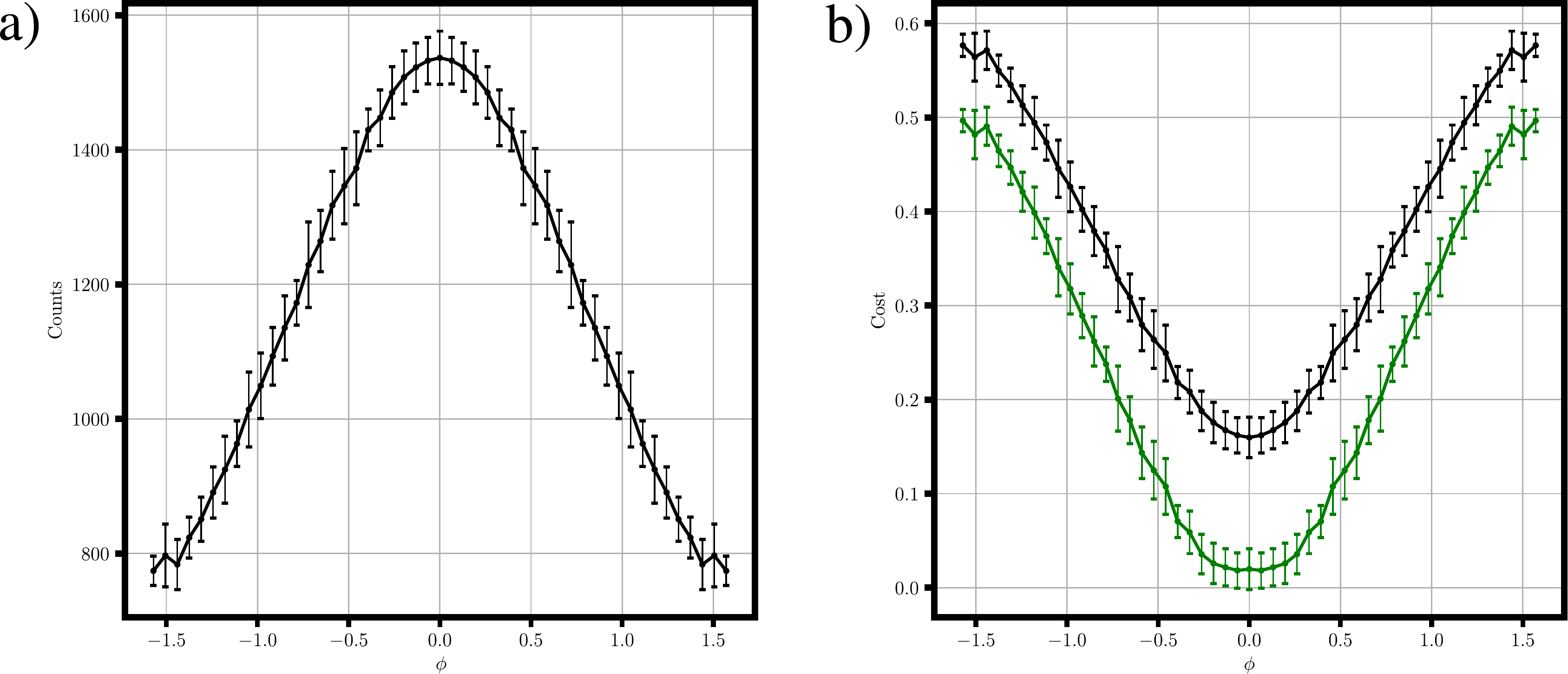}
    \caption{(a) $(0,1,0,1)$ counts used to calculate $Q((0,1,0,1)_{0145},\phi)$. 20 runs of $5\times 10^{4}$ shots per angle $\phi$. (b) (Black) cost function estimate (Eq.(\ref{eqn:costest}) with $P((0,1,0,1)_{0145})$ instead of $P((0,1,0,1)_{2367})$) with numerator $Q((0,1,0,1)_{0145},\phi)$ computed with 20 runs of $5\times 10^{4}$ shots per angle $\phi$. Error bars refer to error in $Q((0,1,0,1)_{0145},\phi)$. Denominator $P((0,1,0,1)_{0145})$ is calculated from 20 runs of $5\times 10^{4}$ shots (average $1828.75$, Bessel corrected standard deviation estimator $48.36$).  (Green) cost function estimate with numerator $Q((0,1,0,1)_{0145},\phi)$ the same as for the black curve, but with regularized denominator given by noisy simulation.}
    \label{fig:twtwtw}
\end{figure*}

The default configuration of the X8 (viz., with no circuit submitted to the device) consists of a network of \texttt{MZgate} and \texttt{Rgate} that compile to the identity (this motivates the consideration that the linear optical circuit in the X8 is based on the rectangular decomposition \cite{rect}). The \texttt{Rgate} is a phase space rotation on a given mode, whereas \texttt{MZgate} is a two-mode linear optical unitary parametrized by two angles $\phi_{1}$, $\phi_{2}$ via
\begin{equation}
U_{\text{MZ}}(\phi_{1},\phi_{2})=U_{\text{BS}}({\pi\over 4},{\pi \over 2})e^{i\phi_{1}a^{\dagger}a} U_{\text{BS}}({\pi\over 4},{\pi \over 2})e^{i\phi_{2}a^{\dagger}a}.
\end{equation} When a four-mode linear optical operation is specified on the X8, it is compiled to the simplest word from the \texttt{MZgate} and \texttt{Rgate} alphabet using the rectangular decomposition \cite{rect}. To minimize loss incurred from inexact parameters produced by this native compilation, it is advantageous to calculate the rectangular decomposition independently, and load the resulting circuit of \texttt{MZgate} and \texttt{Rgate} onto the X8.  For Example 2, one would use
\begin{equation}
U_{\text{BS}}(\theta,\varphi)=e^{i(\varphi+\pi)a^{\dagger}a}U_{\text{MZ}}(\pi -2\theta,2\pi-\varphi)
\label{eqn:rectmz}
\end{equation} 
up to a factor of a complex number of modulus 1. For a given number of experiment runs and shots per angle, we found that specifying (\ref{eqn:rectmz}) as the circuit sent to the X8 gave a smoother cost function  estimate $\hat{D_{2}}(\phi)$ than sending the parametrized $\texttt{BSgate}$. 

Since $P$ does not depend on $\phi$, the ratio $\hat{Q}\over \hat{P}$ can be obtained by dividing the number of $(0,1,0,1)$ counts at $\phi$ and dividing by the number of $(0,1,0,1)$ from a system with the same two-mode squeezed input, but no beamsplitter (the same number of shots should be used to obtain both counts). The latter number can be computed separately one time to best possible precision. This procedure works both for numerical simulations and for X8 runs. We first calculate $Q$ for each angle $\phi$ on the X8. The result for the $(0,1,0,1)$ count is shown in Fig. \ref{fig:twtwtw}a.   At $\phi=0$, the X8 gives an empirical count of Fock state $\ket{0,1,0,1}$ on register 0145 at a lower rate than the rate ${\tanh^{2}1\over \cosh^{4}1}\approx 0.1$ expected from analytical calculation. Near $\phi=\pm {\pi\over 2}$, the X8 gives a higher empirical count of the same Fock state than the analytical result $0$. We observed that loading vacuum as the input state to X8 and running through the circuit $V({\pi\over 4},\phi)_{01}V({\pi\over 4},\phi)_{45}$ specified by \texttt{Rgate} and \texttt{MZgate} as in (\ref{eqn:rectmz}) gives an empirical $\ket{0,1,0,1}$ count of about $0.01$\% at $\phi\approx \pm {\pi\over 2}$ and about $0.1$\% at $\phi \approx 0$; these values are within the error bars of Fig. \ref{fig:twtwtw}a. To better understand the counts in Fig.\ref{fig:twtwtw}a, it would be ideal to have a numerical simulation that qualitatively reproduces the counts for all $\phi$. Unfortunately, we were was unable to determine a composition of attenuation channels \cite{holevoprob} (available through the StrawberryFields function \texttt{ThermalLossChannel} with variable tranmissivity and thermal environment energy parameters) that globally reproduce the $(0,1,0,1)$ counts in Fig.\ref{fig:twtwtw}a.  For example, at $\phi=0$, applying \texttt{ThermalLossChannel} with transmissivity $0.9$ and thermal environment of $1.5$ photons either before or after the  beamsplitter $V({\pi\over 4},\phi=0)$ reduces the $(0,1,0,1)$ count from the analytically expected $(50,000)\cdot{\tanh^{2}1\over \cosh^{4}1} \approx 5,115$ to the range 1,500-2,000, which is closer to the X8 count. However, the agreement is not global over $\phi$ in either case. The low accuracy of an estimate of the probability of $(0,1,0,1)$ from data in Fig. \ref{fig:twtwtw}a is likely due to a combination of photon detection inefficiency \cite{arra}, thermal loss in the linear optical circuit, and squeezing imperfections.

To produce the cost function estimate $\hat{D_{2}}(\phi)$ in the black line of Fig. \ref{fig:twtwtw}b, the the count data in Fig. \ref{fig:twtwtw}a was combined with the estimate of $P((0,1,0,1)_{0145})$ on the X8 according to (\ref{eqn:costest}). For 20 runs of $5\times 10^{4}$ shots, the $(0,1,0,1)_{0145}$ count of 1828.75$\pm$48 was obtained on the X8. For $5\times 10^{4}$ shots ($10^{5}$ shots), the $(0,1,0,1)_{0145}$ count of 1538 (3101) was obtained from numerical simulation with \texttt{ThermalLossChannel}$(0.9,2.0)$. This value was obtained using a Fock basis measurement after dynamics are calculated with the Gaussian backend of StrawberryFields.  If the X8 estimate of $P((0,1,0,1)_{0145})$ is replaced by the value 1538 from the noisy simulation, but the X8 estimate of $Q((0,1,0,1)_{0145},\phi)$ is carried over, the result is a regularized cost function estimate shown by the green lines in Fig. \ref{fig:twtwtw}. This regularized function obtains a value much closer to the analytical value of $0$ at $\phi=0$.

Despite the fact that the cost function estimate in Fig. \ref{fig:twtwtw} does not reach the analytical global minimum, the global critical point $\phi_{c}=0$ can be identified on the $\phi$ grid used. Up to an uncertainty imposed by this discretization of the parameter manifold, this experiment demonstrates strong optimal parameter resilience \cite{opr} of the cost function (\ref{eqn:truecost}) to noise from the X8. A full variational quantum compiling algorithm for a two-mode linear optical unitary may not yet be a convenient implementation of the X8 device, but this possibility is not limited by low-quality cost function estimation.

The conclusion obtained from calculating the cost function (\ref{eqn:costest})  raises many questions about compiling general linear optical unitaries on photonic hardware. For instance, we have taken the ansatz for the beamsplitter to have depth 2 because we know the precompiling scheme for the X8. For a totally different architecture with depth $2L$ and a different, unknown precompiling scheme, one could consider specifying an ansatz of the form
\begin{equation}
\prod_{j=1}^{L}e^{-i\phi_{j}a_{A_{1}}^{\dagger}a_{A_{1}}}U_{\text{BS}}(\xi_{j},0)_{A_{1}A_{2}}
\label{eqn:lay}
\end{equation}
with fixed $\sum_{j=1}^{L}\xi_{j}$, say, $\sum_{j=1}^{L}\xi_{j}={\pi\over 4}$. Expression (\ref{eqn:lay}) is the real beamsplitter $U_{\text{BS}}({\pi\over 4},0)$ iff $\sum_{j=1}^{L}\phi_{j}=0$. Depending on the photon losses in each layer of the hypothetical device and the quality of the precompiling, a greater sample complexity could be required for estimating the cost function using this ansatz (\ref{eqn:lay}). This example emphasizes the importance of accessibility of device specifications to end users.

Finally, we briefly comment on a possible classical module of the variational compiling algorithm, which we assume to be based on applying gradient descent utilizing an estimate of the derivative of cost function. The derivative of $q((0,1,0,1)_{0145},\phi)$ in (\ref{eqn:truecost}) or the numerator of (\ref{eqn:tr}) with respect to $\phi$ can be computed using a CV parameter shift rule of \cite{PhysRevA.99.032331} which utilizes beamsplitters with the same transmissivity and with $\phi$ shifted by $\pm \pi/2$. A full variational quantum compiling algorithm would use a classical module commanding the X8 to locally compute these derivatives. The values of the derivatives would be used to update $\phi$ for the next step of the optimization.

\section*{Acknowledgments}
\pst
The author thanks J.M. Arrazola, N. Killoran, C. Weedbrook, A. Sornborger, and Z. Holmes for helpful conversations, and C. Albornoz, T. Bromley and D. Mahler for discussions at \texttt{discuss.pennylane.ai}. The author acknowledges support from the LDRD program at Los Alamos National Laboratory.  Los Alamos National Laboratory is managed by Triad National Security, LLC, for the National Nuclear Security Administration of the U.S. Department of Energy under Contract No. 89233218CNA000001.

\bibliographystyle{unsrt_style}
\bibliography{tw}

\end{document}